\documentclass [10pt]{article}
\usepackage {graphicx}

\begin{document}

\title{\textbf{ANALYTICAL DESCRIPTION OF HADRON-HADRON SCATTERING VIA PRINCIPLE OF
MINIMUM DISTANCE IN SPACE OF STATES}}
\author{D. B. Ion$^{1,2)}$ and M. L. D. Ion$^{3)}$ \\
$^{1)}$TH Division, CERN, CH-1211 Geneva 23, Switzerland \\
$^{2)}$NIPNE-HH , Bucharest P.O Box MG-6. Romania\\
$^{3)}$Bucharest University, Department of Atomic \\ and Nuclear Physics,
Bucharest, Romania}
\date{}
\maketitle

\begin{abstract}
In this paper an analytical
description of the hadron-hadron scattering is presented by using
PMD-SQS-optimum principle in which the differential cross sections in the
forward and backward c.m. angles are considered fixed from the experimental
data. Experimental tests of the PMD-SQS-optimal predictions, obained by
using the available phase shifts, as well as from direct experimental data,
are presented. It is shown that the actual experimental data for the
differential cross sections of all principal hadron-hadron [nucleon-nucleon,
antiproton-proton, mezon-nucleon] scatterings at all energies higher than 2
GeV, can be well systematized by PMD-SQS predictions.
\end{abstract}

\begin{center}
\textbf{Introduction}
\end{center}

The mathematician Leonhard Euler (1707-1783) appears to have been a
philosophical optimist having written:

\textit{"For since the fabric of the universe is most perfect and
the work of a most wise Creator, nothing at all takes place in the universe
in which some rule of maximum or minimum does not appear. Wherefore, there
is absolutely no doubt that every effect in universe can be explained as
satisfactory from final causes themselves the aid of the method of Maxima
and Minima, as can from the effective causes"}.

Yet this brilliant idea produced many strikingly simple formulations of
certain complex laws of nature. From historical point of view the earliest
optimum principle was proposed by Heron of Alexandria (125 B.C.) in
connection with the behavior of light. Thus, Heron proved mathematically
the following first genuine scientific minimum principle of physics: that
\textit{light travels between two points by shortest path}. In
fact the Archimedean definition of a straight line as the shortest path
between two points was an early expression of a variational principle,
leading to the modern idea of a geodesic path. In fact in the same spirit,
Hero of Alexandria explained the paths of reflected rays of light based on
the principle of minimum distance (PMD), which Fermat (1657) reinterpreted
as a principle of least time, Subsequently, Maupertuis and others developed
this approach into a general principle of least action, applicable to
mechanical as well as to optical phenomena. Of course, a more correct
statement of these optimum principles is that systems evolve along
stationary paths, which may be maximal, minimal, or neither (at an
inflection point). Laws of mechanics were first formulated in terms of
minimum principles. Optics and mechanics were brought together by a single
minimum principle conceived by W. R. Hamilton. From Hamilton's single
minimum principle could be obtained all the optical and mechanical laws then
known. But the effort to find optimum principles has not been confined
entirely to the exact sciences. In modern time the principles of optimum are
extended to all sciences. So, there exists many minimum principle in action
in all sciences, such as: principle of minimum action, principle of
minimum free-energy, minimum charge, minimum entropy production, minimum
Fischer information, minimum potential energy, minimum rate of energy
dissipation, minimum dissipation, minimum of Chemical distance, minimum cross
entropy, minimum complexity in evolution, minimum frustration, minimum sensitivity, etc.
So, a variety of generalizations of classical
variational principles have appeared, and we shall not describe them here.

Next, having in mind this kind of optimism in the paper [1-16] we introduced
and investigated the possibility to construct a predictive analytic theory
of the elementary particle interaction based on the principle of minimum
distance in the space of quantum states (PMD-SQS). So, choosing the partial
transition amplitudes as the system variational variables and the
``distance'' in the Hilbert space of the quantum transitions as a measure of
the system effectiveness expressed in function of partial transition
amplitudes we obtained the results [1-16]. These results proved that the
principle of minimum distance in space of quantum states (PMD-SQS) can be
chosen as variational principle by which we can find the analytic
expressions of the partial transition amplitudes In this project by using
the S-matrix theory the minimum principle PMD-SQS will be formulated in a
general mathematical form. We prove that the new analytic theory of the
quantum physics based on PMD-SQS is completely described with the aid of the
reproducing kernels from RKHS of the transition amplitudes. [1-5].

Therefore, in Ref. [1] by using reproducing kernel Hilbert space (RKHS)
methods [3-5,17], we described the quantum scattering of the spinless
particles by a \textit{principle of minimum distance in the space of quantum states}
(PMD-SQS). Some preliminary experimental tests of the
PMD-SQS, even in the crude form [1] when the complications due to the
particle spins are neglected, showed that the actual experimental data for
the differential cross sections of all principal hadron-hadron
[nucleon-nucleon, antiproton-proton, mezon-nucleon] scatterings at all
energies higher than 2 GeV, can be well systematized by PMD-SQS predictions
(see the papers [1,], Moreover, connections between the PMD-SQS and the
\textit{maximum entropy principle} for the statistics of the scattering quantum channels was also recently
established by introducing quantum scattering entropies: S$_\theta $and
S$_{J }$[5-7]. Then, it was shown that the experimental pion-nucleon as well
as pion-nucleus scattering entropies are well described by optimal entropies
and that the experimental data are consistent with the principle of minimum
distance in the space of quantum states (PMD-SQS) {\}}[1]. However, the
PMD-SQS in the crude form [1] cannot describe the polarization J-spin
effects.

In this paper an analytical description of the hadron-hadron scattering is
presented by using PMD-SQS-optimum principle in which the differential cross
sections in the forward (x=+1) and backward (x=-1) directions are considered
fixed from the experimental data. An experimental test of the optimal
prediction on the logarithmic slope b is performed for the pion-nucleon and
kaon-nucleon scatterings at the forward c.m. angles.

\begin{center}
\textbf{2. Description of pion-nucleon scattering via principle of minimum
distance in space of quantum states (PMD-SQS)}
\end{center}

First we present the basic definitions on the $(0^ - 1 / 2^ + \to 0^ - 1 /
2^ + )$ hadronic scattering:

\begin{equation}
\label{eq1}
M(0^ - ) + N(1 / 2^ + ) \to M(0^ - ) + N(1 / 2^ + ),
\end{equation}

Therefore, let $f^{ + + }(x)$and $f^{ + - }(x)$, be the scattering helicity
amplitudes of the mezon-nucleon scattering process (see ref.[14]) written in
terms of the partial helicities $f_{J - } $ sand $f_{J + } $as follows

\begin{equation}
\label{eq2}
\begin{array}{l}
 f_{ + + } \left( x \right) = \sum\limits_{J = \frac{1}{2}}^{J_{\max } }
{\left( {J + \frac{1}{2}} \right)} \left( {f_{J - } + f_{J + } }
\right)d_{\frac{1}{2}\frac{1}{2}}^J \left( x \right) \\
 f_{ + - } \left( x \right) = \sum\limits_{J = \frac{1}{2}}^{J_{\max } }
{\left( {J + \frac{1}{2}} \right)} \left( {f_{J - } - f_{J + } } \right)d_{
- \frac{1}{2}\frac{1}{2}}^J \left( x \right) \\
 \end{array}
\end{equation}

\noindent
where the rotation functions are defined as

\begin{equation}
\label{eq3}
\begin{array}{l}
 d_{\frac{1}{2}\frac{1}{2}}^J \left( x \right) = \frac{1}{l + 1} \cdot
\left[ {\frac{1 + x}{2}} \right]^{\frac{1}{2}}\left[ {\mathop
P\limits^\bullet _{l + 1} \left( x \right) - \mathop P\limits^\bullet _l
\left( x \right)} \right] \\
 d_{ - \frac{1}{2}\frac{1}{2}}^J \left( x \right) = \frac{1}{l + 1} \cdot
\left[ {\frac{1 - x}{2}} \right]^{\frac{1}{2}}\left[ {\mathop
P\limits^\bullet _{l + 1} \left( x \right) + \mathop P\limits^\bullet _l
\left( x \right)} \right] \\
 \end{array}
\end{equation}

\noindent
where $P_l (x) $are Legendre polinomials, $\mathop P\limits^o _l (x) =
\frac{d}{dx}P_l (x)$, x being the c.m. scattering angle. The normalisation
of the helicity amplitudes $f^{ + + }(x)$ and $f^{ + - }(x)$, is chosen such
that the c.m. differential cross section is given by

\begin{equation}
\label{eq4}
\frac{d\sigma }{d\Omega }\left( x \right) = \left| f_{ + + } \left( x\right) \right|^2 +
\left| f_{ + - } \left( x \right) \right|^2
\end{equation}

Then, the elastic integrated cross section is given by

\begin{equation}
\label{eq5a} \frac{\sigma_{el} }{2\pi } =
\sum\limits_{J=frac{1}{2}}^{J_max }\left( 2J+1\right)\left( \left| f_{J+}\right|^2 +
\left| f_{J-}\right|^2 \right)
\end{equation}

Now, let us consider the following optimization problem:

\begin{equation}
D(f_{J+},f_{j-})=\sigma _{el} / 2\pi =\sum {(j + \frac{1}{2})\left[ {\left| f_{j+} \right| ^2 +
\left| f_{j-} \right| ^2} \right]}
\end{equation}

\noindent
when $\frac{d\sigma }{d\Omega }( + 1)\,\,\,$and $\frac{d\sigma }{d\Omega }(
- 1)$ are fixed.

We proved that the solution of this optimization problem is given by the
following results :

\begin{equation}
f_{o}^{++}(x)= f^{++}(+1)\frac{K_{\frac{1}{2}\frac{1}{2}}(x,y)}{K_{\frac{1}{2}\frac{1}{2}}(+1,+1)}
\end{equation}

\begin{equation}
f_{o}^{+-}(x)= f^{+-}(-1)\frac{K_{\frac{1}{2}-\frac{1}{2}}(x,y)}{K_{\frac{1}{2}-\frac{1}{2}}(-1,-1)}
\end{equation}

\noindent
where the functions K(x,y) are the \textit{reproducing kernels} [3-5] expressed in terms of rotation
function by

\begin{equation}
K_{\frac{1}{2}\frac{1}{2}} (x,y) = \sum\limits_{1 / 2}^{J_o } ( j +
\frac{1}{2})d_{\frac{1}{2}\frac{1}{2}}^j (x)d_{\frac{1}{2}\frac{1}{2}}^j
(y),
\end{equation}

\begin{equation}
K_{\frac{1}{2} - \frac{1}{2}} (x,y) = \sum\limits_{1 / 2}^{J_o } ( j +
\frac{1}{2})d_{\frac{1}{2} - \frac{1}{2}}^j (x)d_{\frac{1}{2} -
\frac{1}{2}}^j (y)
\end{equation}

\noindent
while the optimal angular momentum is given by

\begin{equation}
\label{eq11}
J_0 = \sqrt {\frac{4\pi }{\sigma _{el} }\left[ {\frac{d\sigma }{d\Omega }( +
1) + \frac{d\sigma }{d\Omega }( - 1)} \right] + \frac{1}{4}} - 1
\end{equation}

Now, let us consider the logarithmic slope b of the forward diffraction peak
defined by

\begin{equation}
b =\frac{d}{dt}\left[ {\ln \frac{d\sigma }{dt}(s,t)} \right]_{t = 0}
\end{equation}

Then, using the definition of the rotation functions, from (7)-(\ref{eq11}) we
obtain the optimal slope $b_{o}$

\begin{equation}
\label{eq13}
b_o = \frac{\lambda ^2}{4}\left[ {\frac{4\pi }{\sigma _{el} }\left(
{\frac{d\sigma }{d\Omega }( + 1) + \frac{d\sigma }{d\Omega }( - 1)} \right)
- 1} \right]
\end{equation}

Finally, we note that in ref. [13] we proved the following optimal
inequality

\begin{equation}
\label{eq14}
b_o = \frac{\lambda ^2}{4}\left[ {\frac{4\pi }{\sigma _{el} }\left(
{\frac{d\sigma }{d\Omega }( + 1) + \frac{d\sigma }{d\Omega }( - 1)} \right)
- 1} \right] \le b_{\exp }
\end{equation}

\noindent
which includes in a more general and exact form the unitarity bounds derived
by Martin [18] and Martin-Mac Dowell [19] (see also ref.[20]) and Ion
[1,21]. Indeed, since $\frac{d\sigma }{d\Omega }(\pm 1) \ge 0$, and

\begin{equation}
\frac{d\sigma }{d\Omega }( + 1) \ge \frac{\sigma _T^2 }{16\pi
\mathchar'26\mkern-10mu\lambda ^2},
\end{equation}
(Wik inequality)
\noindent
from the bound (\ref{eq14}), we get

\begin{equation}
\frac{\lambda ^2}{4}\left[ {\frac{4\pi }{\sigma _{el} }\left( {\frac{d\sigma
}{d\Omega }( + 1)} \right) - 1} \right] \le b_{\exp } \:\rm{(proved\: in\: ref.\: [1])}
\end{equation}

\begin{equation}
\frac{\lambda ^2}{4}\left[ {\frac{4\pi }{\sigma _{el} }\left( {\frac{d\sigma
}{d\Omega }( - 1)} \right) - 1} \right] \le b_{\exp } \:\rm{(proved\: in\: this\: paper)}
\end{equation}

\begin{equation}
\frac{\mathchar'26\mkern-10mu\lambda ^2}{4}\left[ {\frac{\sigma _T^2 }{4\pi
\mathchar'26\mkern-10mu\lambda ^2\sigma _{el} } - 1} \right] \le b_{\exp }
\:\rm{(improved\: Martin-MacDowell\: bound\: [19])}
\end{equation}

\begin{equation}
\frac{\mathchar'26\mkern-10mu\lambda ^2}{4}\left[ {\frac{\sigma _T }{4\pi
\mathchar'26\mkern-10mu\lambda ^2} - 1} \right] \le b_{\exp } \:\rm{(Martin \: bound\:
[18])}
\end{equation}

\begin{center}
\textbf{3. Experimental tests of the PMD-SQS-optimal predictions}
\end{center}

For an experimental test of the optimal result (\ref{eq14}) the numerical values of
the slopes $b_{o }$ and $b_{exp }$ are calculated directly by reconstruction
of the helicity amplitudes from the experimental phase shifts (EPS)
solutions of Holer et al. [23] and also directly from the experimental data.
The results are displayied in Fig 1-5. Moreover, we calculated from the
experimental data (see [24-29]) the following physical quantities:

SCALING FUNCTION:
\begin{equation}
f(\tau ) \equiv \frac{d\sigma }{d\Omega }(x) /
\frac{d\sigma }{d\Omega }(\ref{eq1})
\end{equation}

SCALING VARIABLE:
\begin{equation}
\tau \equiv 2\sqrt {\vert t\vert b_o }
\end{equation}

\noindent
and compared with the values of the PMD-SQS-optimal predictions obtained
from

OPTIMAL SCALING FUNCTION:
\begin{equation}
f^o(\tau _o ) \equiv \frac{d\sigma ^0}{d\Omega
}(x) / \frac{d\sigma }{d\Omega }(\ref{eq1}) = \left[
{\frac{K_{\frac{1}{2}\frac{1}{2}} (x,1)}{K_{\frac{1}{2}\frac{1}{2}} (1,1)}}
\right]^2 \approx \left[ {\frac{2J_1 (\tau _o )}{\tau _o }} \right]^2
\end{equation}

The results are presented in Fig. 6. We must note that the approximation in
(22) is derived by using the relation

\begin{equation}
d_{\mu \nu }^j (x) \approx J_{\vert \mu - \nu \vert } \left[2(j + 1)\sin
\frac{\theta }{2}\right],\: \rm{for\: small\: \theta - angles}
\end{equation}

Where $J_{\vert \mu - \nu \vert } (\tau )$are Bessel functions of order
$\vert \mu - \nu \vert $.

\begin{figure}[htbp]
\centerline{\includegraphics[width=12cm]{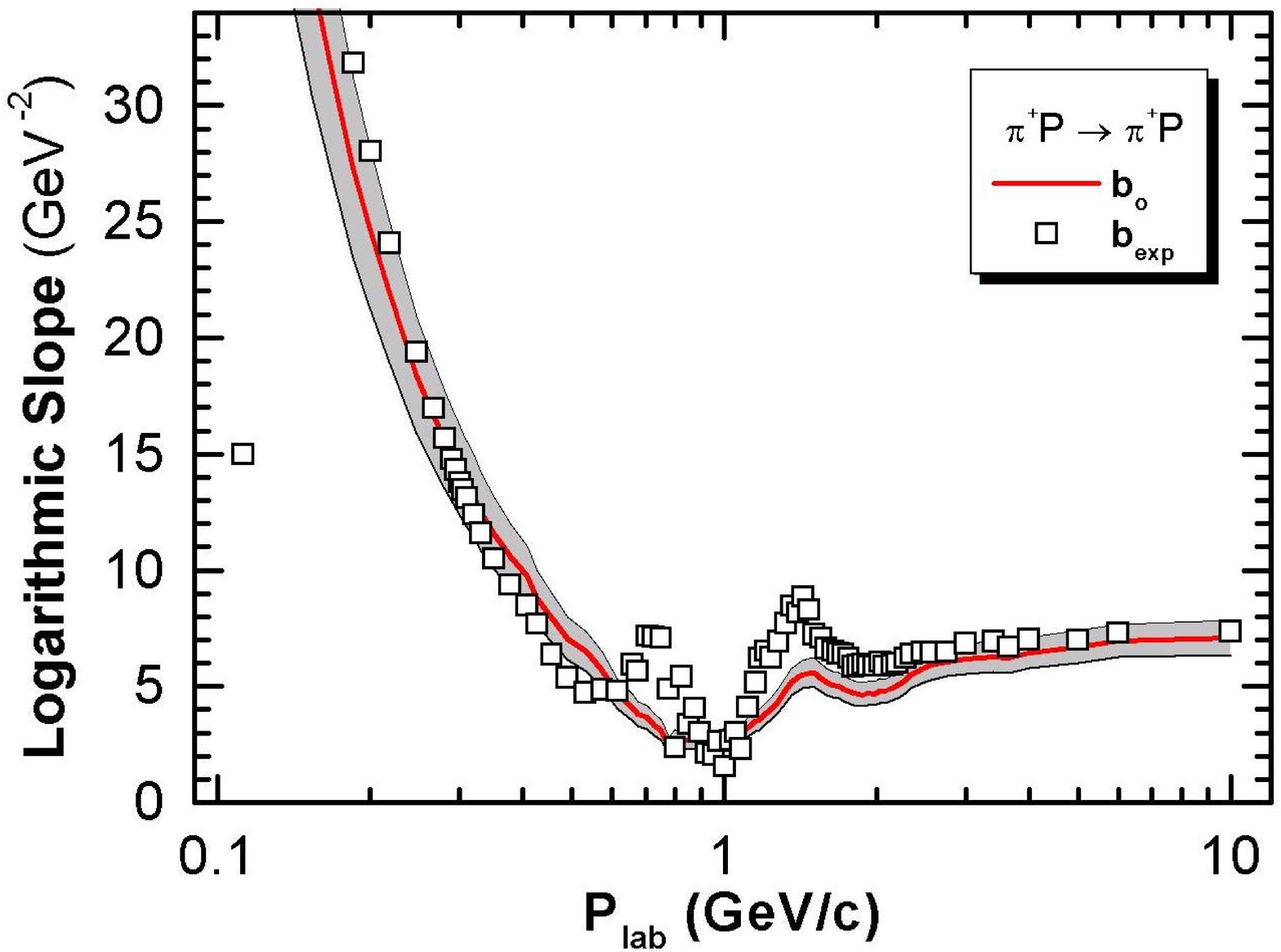}} \label{fig2}
\caption{The experimental logarithmic slopes (b$_{exp})$ of the
diffraction peak, for the forward $\pi ^ + P \to \pi ^ + P$
scattering, are compared with the optimal predictions b$_{o}$
(\ref{eq13}).}
\end{figure}

\begin{figure}[htbp]
\centerline{\includegraphics[width=15cm]{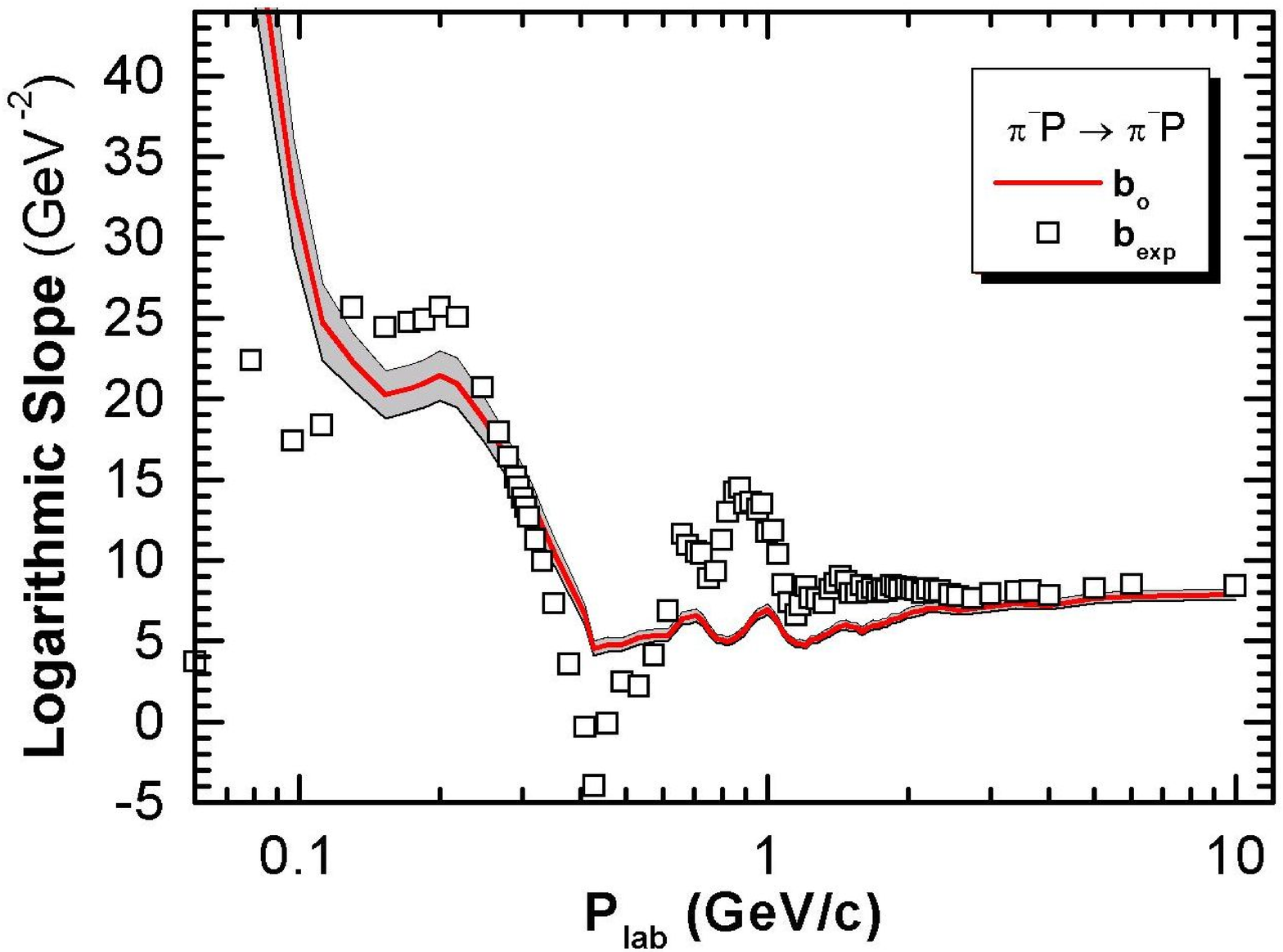}} \label{fig3}
\caption{The experimental logarithmic slopes (b$_{exp})$ of the
diffraction peak, for the forward $\pi ^ = P \to \pi ^ - P$
scattering, are compared with the optimal predictions b$_{o}$
(\ref{eq13}) .}
\end{figure}

\begin{figure}[htbp]
\centerline{\includegraphics[width=15cm]{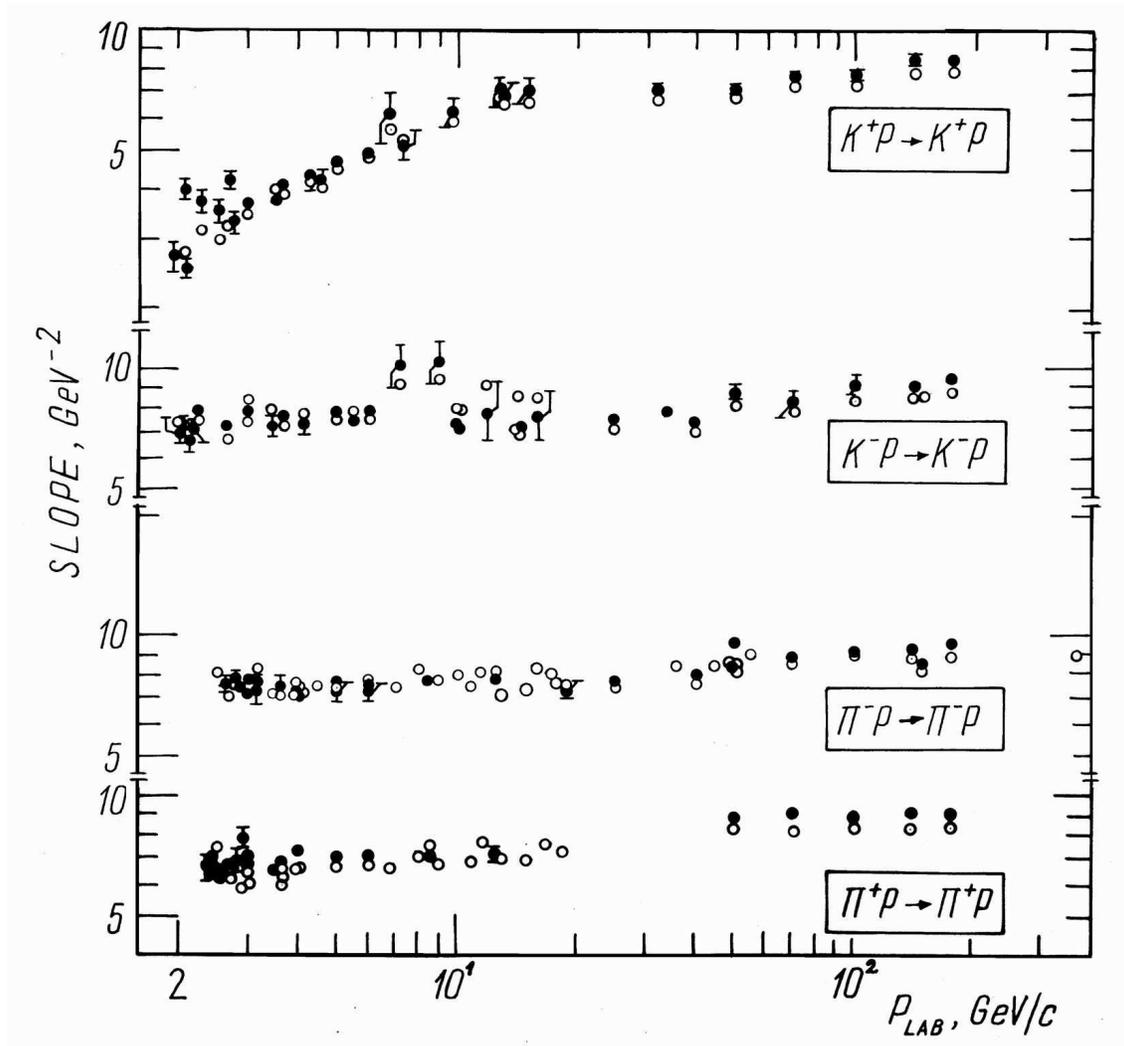}} \caption{The
experimental logarithmic slopes (black circles) of the diffraction
peak, for the forward $\pi ^ = P \to \pi ^ - P$ scattering, are
compared with the optimal predictions b$_{o}$ (\ref{eq13})(white
circles).} \label{fig4}
\end{figure}

\begin{figure}[htbp]
\centerline{\includegraphics[width=15cm]{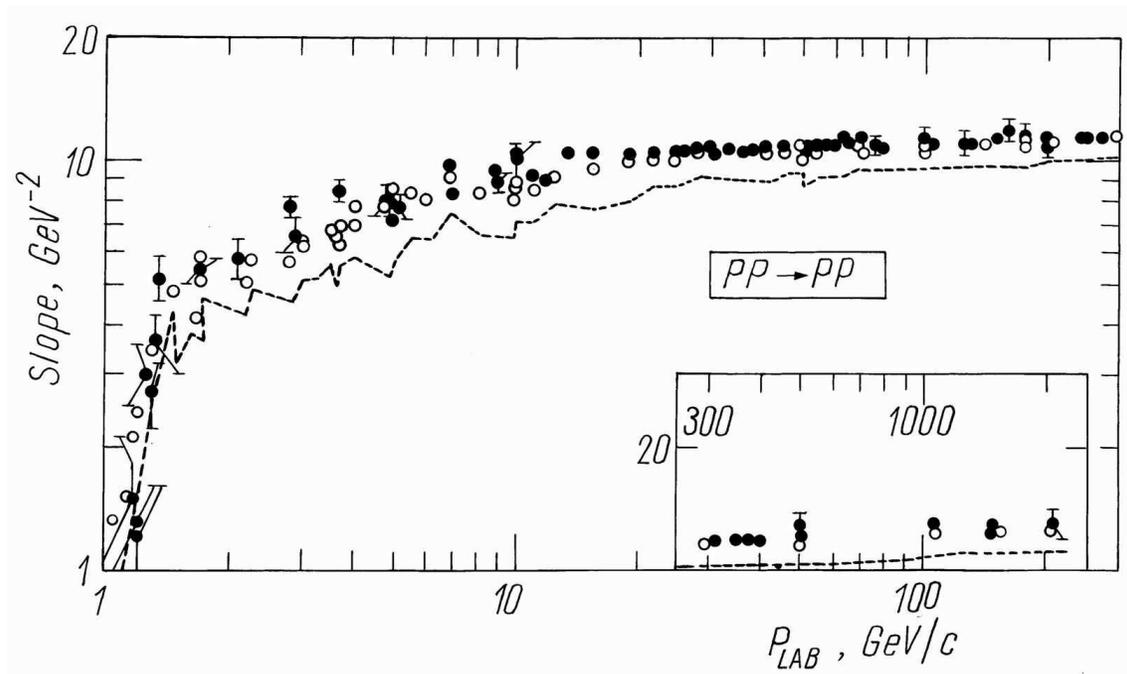}} \label{fig5}
\caption{The experimental values of the logarithmic slopes are
compared with the values of optimal predictions (xx) (solid
curves) for the $PP \to PP$scatterings. Dashed curve correspond to
an estimation of the Martin-MacDowell bound [19].}
\end{figure}

\begin{figure}[htbp]
\centerline{\includegraphics[width=15cm]{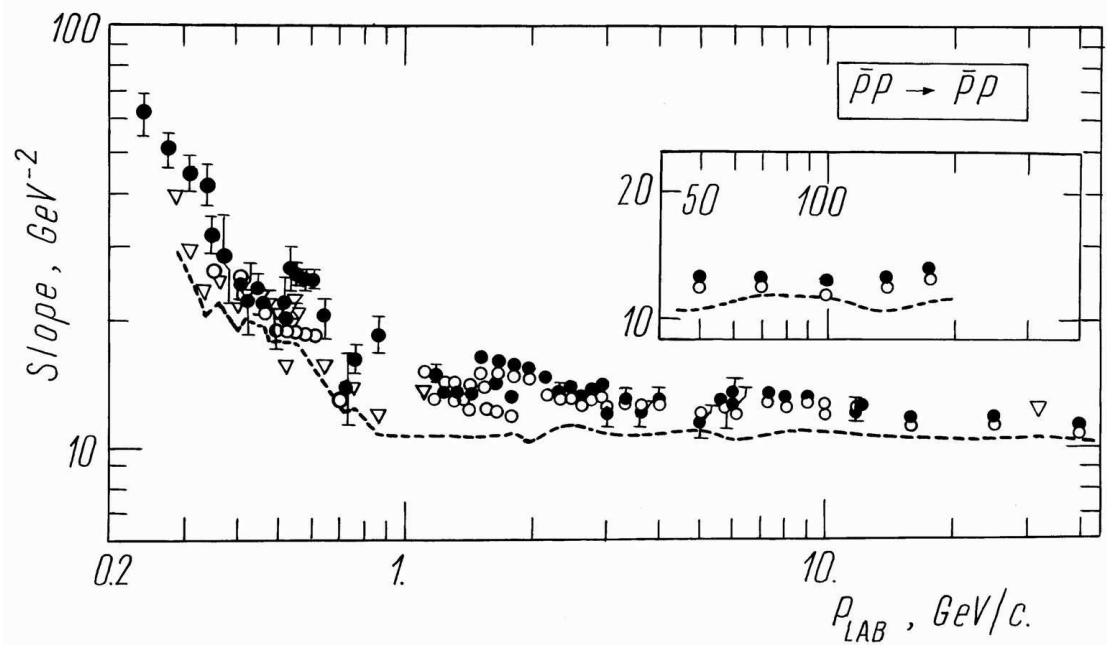}} \label{fig6}
\caption{The experimental values of the logarithmic slopes (black
circles) are compared with the values of optimal predictions
(white circles) for the (a) $\overline P P \to \overline P
P$ scatterings. Dashed curve correspond to an estimation of the
Martin-MacDowell bound [19].}
\end{figure}

\begin{figure}[htbp]
\centerline{\includegraphics[width=15cm]{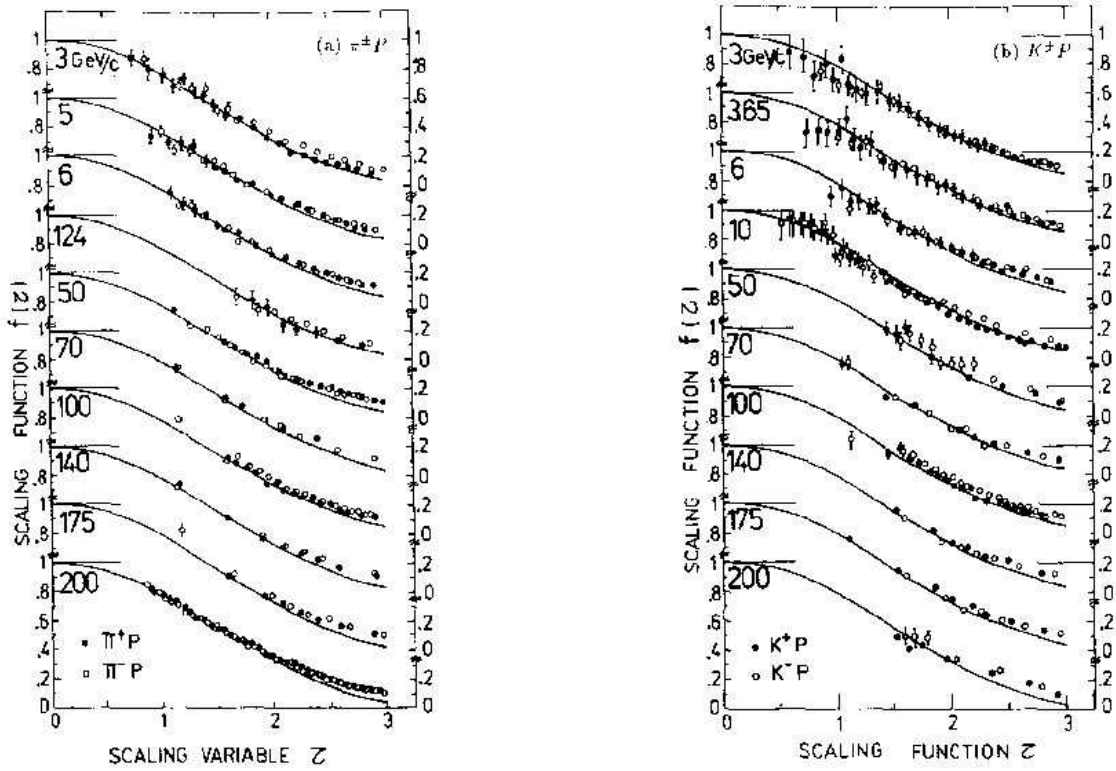}} \label{fig7}
\caption{The experimental values of the scaling function (20) are
compared with the values of optimal scaling predictions (22)
(solid curves) for the (a) $\pi ^\pm P \to \pi ^\pm P$ and (b)
$K^\pm P \to K^\pm P$ scatterings}
\end{figure}

\begin{figure}[htbp]
\centerline{\includegraphics[width=15cm]{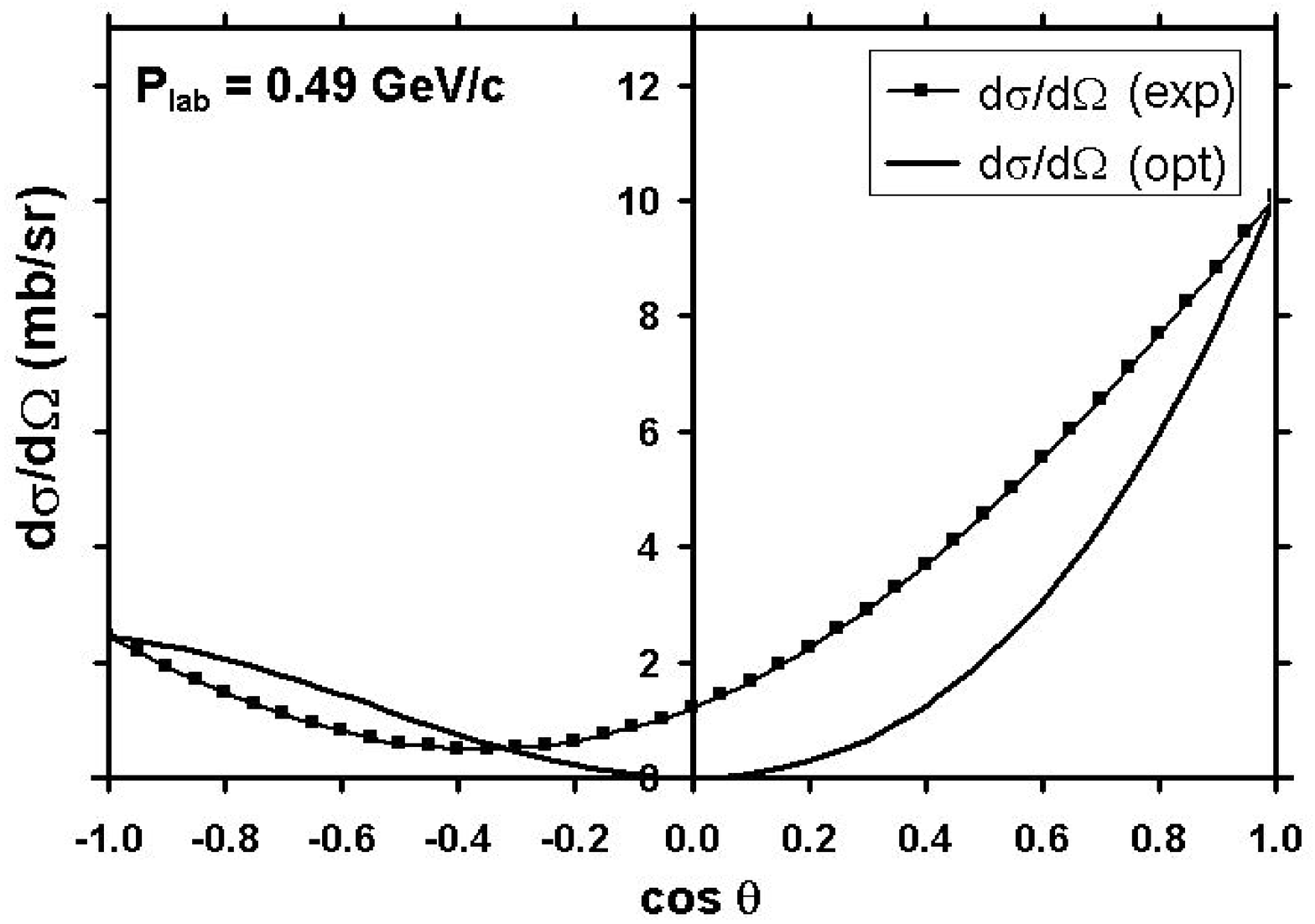}} \label{fig8}
\caption{The differential cross sections for $\pi ^ + P \to \pi ^
+ P$ calculated by using eqs. (\ref{eq4}) and the experimental
phase shifts [12] are compared with the optimal state predictions
given by eqs. (7)-(\ref{eq11}).}
\end{figure}

\begin{figure}[htbp]
\centerline{\includegraphics[width=15cm]{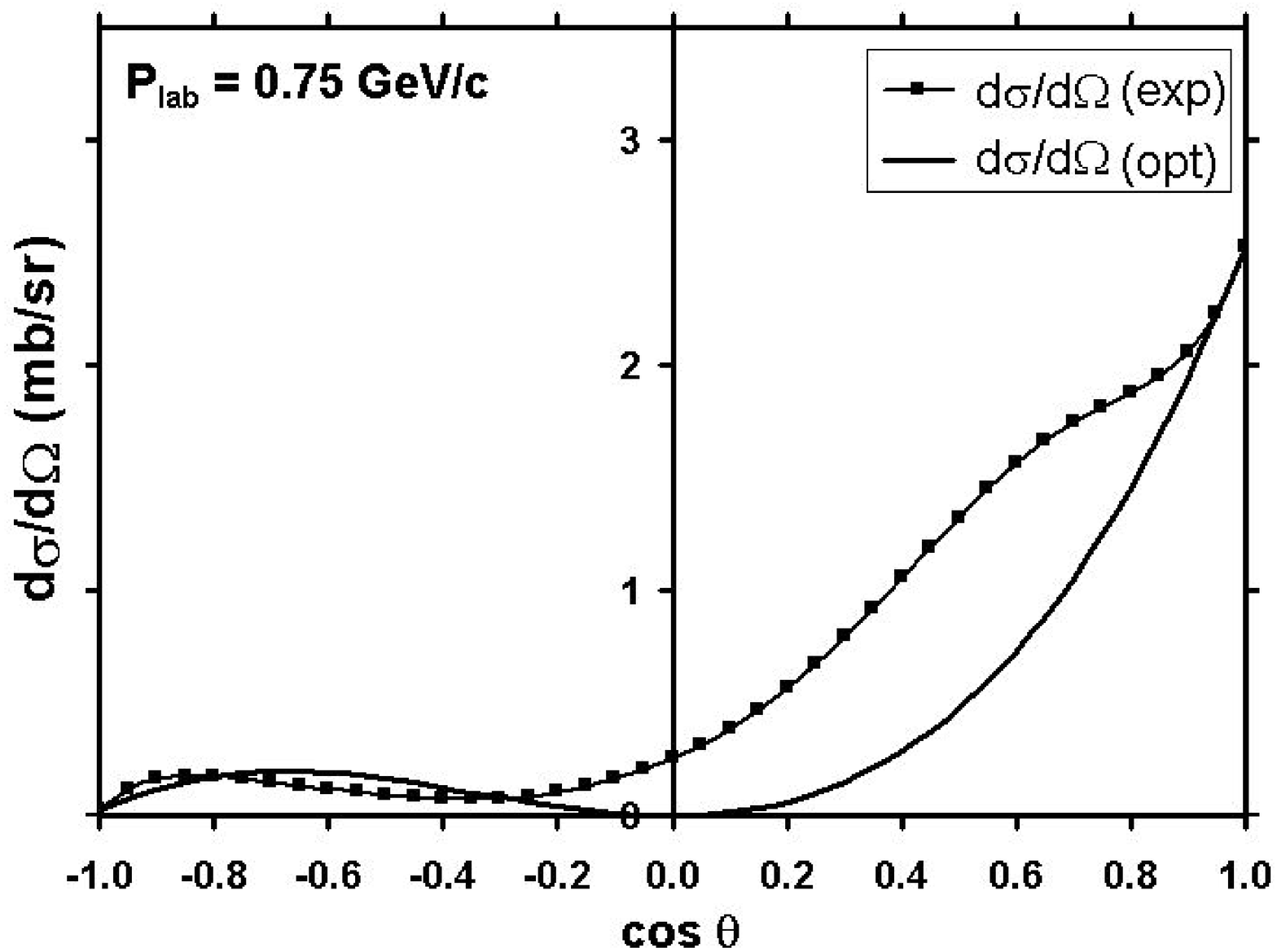}} \label{fig9}
\caption{The differential cross sections for $\pi ^ + P \to \pi ^
+ P$ calculated by using eqs. (\ref{eq4}) and the experimental
phase shifts [12] are compared with the optimal state predictions
given by eqs. (7)-(\ref{eq11}).}
\end{figure}

\begin{figure}[htbp]
\centerline{\includegraphics[width=15cm]{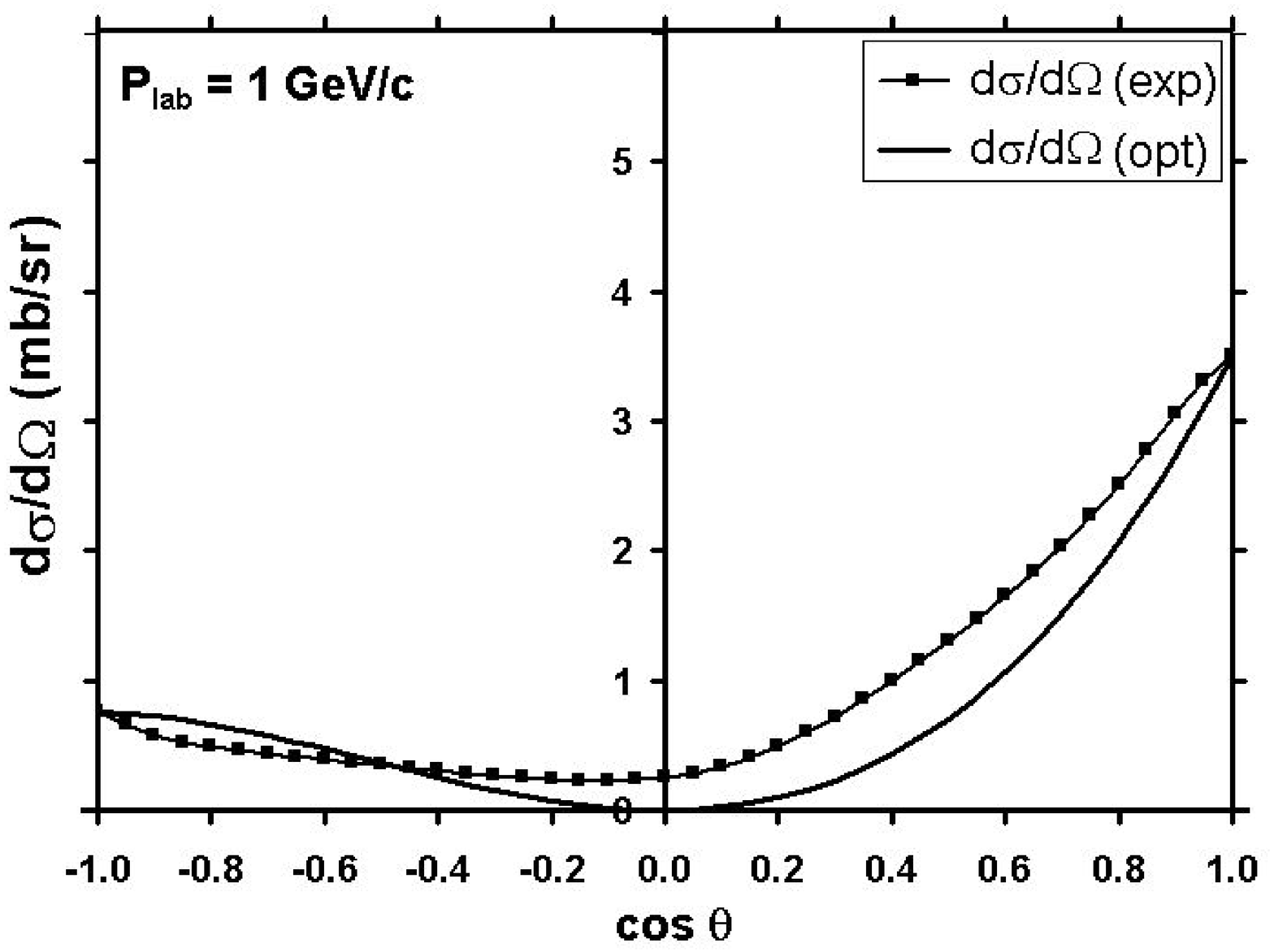}} \label{fig10}
\caption{The differential cross sections for $\pi ^ + P \to \pi ^
+ P$ calculated by using eqs. (\ref{eq4}) and the experimental
phase shifts [12] are compared with the optimal state predictions
given by eqs. (7)-(\ref{eq11}).}
\end{figure}

\begin{figure}[htbp]
\centerline{\includegraphics[width=15cm]{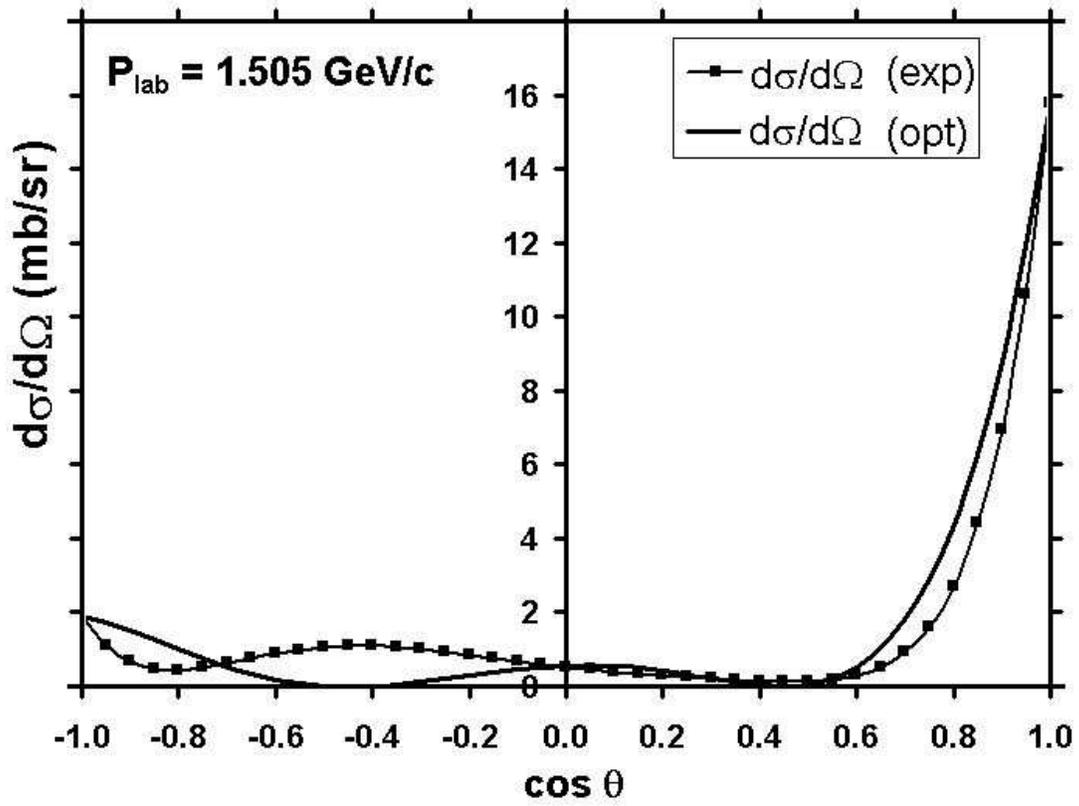}} \label{fig11}
\caption{The differential cross sections for $\pi ^ + P \to \pi ^
+ P$ calculated by using eqs. (\ref{eq4}) and the experimental
phase shifts [12] are compared with the optimal state predictions
given by eqs. (7)-(\ref{eq11}).}
\end{figure}

\begin{center}
\textbf{4. Conclusions }
\end{center}

The main results and conclusions obtained in this paper can be summarized as
follows:

In this paper an analytical description of the hadron-hadron scattering is
presented by using PMD-SQS-optimum principle in which the differential cross
sections in the forward (x=+1) and backward (x=-1) directions are considered
fixed from the experimental data. So, choosing the partial transition
amplitudes as the system variational variables and the ``distance'' in the
Hilbert space of the quantum transitions as a measure of the system
effectiveness expressed in function of partial transition amplitudes we
obtained the results [1-16].

(i) The PMD-SQS optimal dominance in hadron-hadron scattering at small
transfer momenta for $p_{LAB} >2$ GeV/c is a fact well evidenced
experimentally by the results presented in Figs. 1-6. This conclusion can be
also extended in low energy region.

(ii) In the low energy region, the optimal slope (\ref{eq13}) is in good agreement
with the experimental data in some domains of energy between the resonances
positions or/and in the region corresponding to the diffractive resonances
see Figs. 1-2 and Figs. 7-10.

(iii) We find that the presented experimental tests prove that the principle
of minimum distance in space of quantum states (PMD-SQS) can be chosen as
variational principle by which we can find the analytic expressions of the
partial transition amplitudes.

Finally, we hope that our results are encouraging for an analytic
description of the quantum scattering in terms of an optimum principle,
namely, the \textit{principle of minimum distance in space of quantum state} (PMD-SQS) introduced by us in ref. [1].

\begin{center}
\textbf{5. References}
\end{center}

[1] {D.B.Ion} \textit{''Description of quantum scattering via principle of minimum distance
in space of states''}, Phys. Lett. \textbf{B 376} (1996) 282.

[2] {D.B.Ion} and
{M.L.D.Ion},\textit{''Isospin quantum distances in hadron-hadron scatterings''}, Phys. Lett.
\textbf{B 379} (1996) 225.

[3] {D.B.Ion} and H. Scutaru,
\textit{''Reproducing kernel Hilbert space and optimal state description of
hadron-hadron scattering''}, Int. J. Theor. Phys. \textbf{24} (1985) 355.

[4] {D.B.Ion, }\textit{''Reproducing kernel Hilbert spaces and
extremal problems for scattering of particles with arbitrary spins''},
Int. J. Theor. Phys. \textbf{24} (1985)
1217.

[5] {D.B.Ion} \textit{''Scaling and S-channel helicity conservation via optimal
state description of hadron-hadron scattering''}, Int. J. Theor. Phys. \textbf{25} (1986)
1257.

[6] {D.B.Ion }and {M.L.D.Ion}\textit{, ''Information entropies in pion-nucleon
scattering and optimal state analysis''},

Phys. Lett. \textbf{B 352} (1995) 155.

[7] {D.B.Ion }and {M.L.D.Ion},
\textit{''Entropic lower bound for quantum scattering of
spineless particles''}, Phys. Rev. Lett. \textbf{81} (1998) 5714.

[8] M.L.D.Ion and D.B.Ion,
\textit{''Entropic uncertainty relations for nonextensive quantum scattering, }

Phys. Lett. \textbf{B 466} (1999) 27-32.

[9] M.L.D.Ion and D.B.Ion, \textit{ ''Optimal bounds for Tsallis-like entropies
in quantum scattering of spinless particles''}, Phys.
Rev. Lett. \textbf{83} (1999) 463.

[10] M.L.D.Ion and D.B.Ion, \textit{ ''Angle-angular-momentum entropic bounds
and optimal entropies for quantum scattering of spineless particles''}, Phys. Rev.
\textbf{E 60} (1999) 5261.

[11] D. B. Ion and M. L.D. Ion, \textit{''Limited entropic uncertainty
as a new principle in quantum physics'',}
Phys. Lett. \textbf{B 474} (2000) 395.

[12] M.L.D.Ion and D.B.Ion, \textit{ ''Strong evidences for correlated nonextensive
statistics in hadronic scatterings'',} Phys. Lett. \textbf{B 482} (2000) 57.

[13] D. B. Ion and M. L.D. Ion, \textit{''Optimality entropy and complexity in quantum scattering''},

Chaos Solitons and Fractals, \textbf{13} (2002) 547.

[14] D. B. Ion and M. L.D.Ion, \textit{''Evidences for nonextensive statistics
conjugation in hadronic scatterings systems''}, Phys. Lett. \textbf{B 503} (2001) 263.

[15] D. B. Ion and M. L. D.Ion, \textit{''New nonextensive quantum entropy and
strong evidences for the equilibrium of quantum hadronic states}'', Phys. Lett.
\textbf{B 519 } (2001) 63.

[16] D. B. Ion and M.D. Ion\textit{, Nonextensive statistics ans saturation of
PMD-SQS-optimality limits in hadronic scattering'',} Physica \textbf{A 340} (2004)
501.

[17] N. Aronsjain, Proc. Cambridge Philos. Soc. \textbf{39} (1943) 133,
Trans. Amer. Math. Soc. \textbf{68} (1950) 337; S. Bergman, The Kernel
Function and Conformal mapping, Math. Surveys No 5. AMS, Providence, Rhode
Island, 1950; S. Bergman,and M. Schiffer Kernel Functions and Eliptic
Differential Equations in Mathematical Physics, Academic Press, New York,
1953; A. Meschkowski, Hilbertische Raume mit Kernfunction, Springer Berlin,
1962; H.S. Shapiro, Topics in Approximation Theory, Lectures Notes in
Mathematics, No 187, Ch. 6, Springer, Berlin, 1971

[18] A. Martin, Phys. Rev. \textbf{129},1432 (1963).

[19] S. W. MacDowell and A. Martin, Phys. Rev\textbf{. 135 B}, 960 (1964).

[20] S. M. Roy, Phys. Rep. \textbf{5C}, 125 (1972).

[21] D. B. Ion, St. Cerc. Fiz. \textbf{43}, 5 (1991).

[22] See the books: N. R. Hestenes, \textit{Calculus of variations and
optimal control theory}, John Wiley{\&}Sons, Inc., 1966, and also
V. M. Alecseev, V. M. Tihomirov and S. V. Fomin, \textit{Optimalinoe Upravlenie},
Nauka, Moskow,1979.

[23] Hohler, F. Kaiser, R. Koch, E. Pitarinen, Physics Data, Handbook of
Pion Nucleon Scattering, 1979, Nr. 12-1.

[24] D. B. Ion and M. L. Ion, \textit{A new optimal bound on logarithmic slope
of elastic hadro-hadron scattering}, ArXiv: hep-ph/0501146 v1 16 Jan 2005.

[25] For extensive experimental literature see: J. Bystricky
--Landolt-Bornstein, New Series, Group I-Vol 9a, \textit{Nucleon-Nucleon and
Kaon-Nucleon Scattering} (1980).

[26] D.B.Ion, C. Petrascu, Rom. Journ. Phys. \textbf{37} (1992) 569-575.

[27] D.B.Ion, C. Petrascu and A. Rosca, Rom. Journ. Phys. \textbf{37} (1992)
977-989.

[28] D.B.Ion, C. Petrascu, Rom. Journ. Phys. \textbf{38} (1993) 23-29.

[29] D.B.Ion, C. Petrascu, A. Rosca and V. Topor, Rom. Journ. Phys.
\textbf{39} (1994) 213-221.

\end{document}